# Catalytically Potent and Selective Clusterzymes for Modulation of Neuroinflammation Through Single-Atom Substitutions

*Haile Liu[1,†], Yonghui Li[1,†], Si Sun[1,†], Qi Xin[1], Shuhu Liu[3], Xiaoyu Mu[1], Xun Yuan[4], Ke Chen[1], Hao Wang[1], Kalman Varga[5], Wenbo Mi[1], Jiang Yang[6], Xiao-Dong Zhang [1,2] \**

[1] Department of Physics and Tianjin Key Laboratory of Low Dimensional Materials Physics and Preparing Technology, School of Sciences, Tianjin University, Tianjin 300350, China.

[2] Tianjin Key Laboratory of Brain Science and Neural Engineering, Academy of Medical Engineering and Translational Medicine, Tianjin University, Tianjin 300072, China

[3] Beijing Synchrotron Radiation Facility (BSRF), Institute of High Energy Physics (IHEP), Chinese Academy of Sciences (CAS), Beijing, 100049 P. R. China

[4] School of Materials Science and Engineering, Qingdao University of Science and Technology, Qingdao, Shandong 266042, China

[5] Department of Physics and Astronomy, Vanderbilt University, Nashville, Tennessee 37235, USA

[6] School of Medicine, Sun Yat-sen University, Guangzhou 510060, China

*Correspondence to Xiaodong Zhang, Email: xiaodongzhang@tju.edu.cn.

†: These authors have contributed equally.

**ABSTRACT**. Emerging artificial enzymes with reprogrammed and augmented catalytic activity and substrate selectivity have long been pursued with sustained efforts. The majority of current candidates rely on noble metals or transition metal oxides with rather poor catalytic activity compared with natural molecules. To tackle this limitation, we strategically designed a novel artificial enzyme based on a structurally well-defined $Au_{25}$ cluster, namely clusterzyme, which is endowed with intrinsic high catalytic activity and selectivity driven by single-atom substitutions with modulated bond lengths. The 3-mercaptopropionic acid (MPA)-stabilized $Au_{24}Cu_1$ and $Au_{24}Cd_1$ clusterzymes exhibit 137 and 160 times higher antioxidant capacities than the natural trolox, respectively. Meanwhile, the clusterzymes each demonstrate preferential enzyme-mimicking catalytic activities with compelling selectivity: $Au_{25}$ exhibits superior glutathione peroxidase-like (GPx-like) activity; $Au_{24}Cu_1$ shows a distinct advantage towards catalase-like (CAT-like) activity by its Cu single active site; $Au_{24}Cd_1$ preferably acts as a superoxide dismutase-like (SOD-like) enzyme *via* the Cd single active site. This unique diversified catalytic landscape manifests distinctive reactions against inflammation in brain. $Au_{24}Cu_1$ behaves as an endogenous multi-enzyme mimic that directly decreases peroxide in injured brain *via* catalytic reactions, while $Au_{24}Cd_1$, catalyzes superoxide and nitrogenous signal molecules by preference, and significantly decreases inflammation factors such as IL-1β, IL-6, and TNFα, indicative of an important role in mitigating neuroinflammation.

## 1. Introduction

Due to their exclusive catalytic activity and selectivity, artificial enzymes are exploited as promising tools for wide-reaching biomedical implications,[1-8] particularly as advanced diagnostics[9,10] and therapeutics[11-16] of diseases. Earlier studies shed light on the oxidase- and peroxidase-like activities of noble metals.[17] Gold-based materials were unraveled to possess versatile enzyme-like activities such as nuclease, glucose oxidase (GOD), peroxidase (POD), catalase (CAT), and superoxide dismutase (SOD).[17,18] The Michaelis-Menten constant ($K_m$) to the $H_2O_2$ substrate of gold nanoparticles towards the POD enzymatic reaction is below 1 mM, but the catalytic activity is weak.[19] In contrast, Pt-based materials generally confer a high overall catalytic activity but it can only show a good $H_2O_2$ substrate affinity when $K_m$ is up to 16.7 mM,[14,20] and modulation of selective catalysis often needs to be purposely realized through rationally designed combination with other catalysts.[21] Meanwhile, metal oxides have also revealed great potentials as enzyme mimetics.[22] Typically, $Fe_3O_4$ nanoparticles display the POD-like activity,[23,24] but are limited by their affinity to the $H_2O_2$ substrate ($K_m$ at ~ 154 mM) and a maximal reaction rate (5.9 μM/min) that do not meet expectations. $Mn_3O_4$ nanoparticles concurrently exhibit SOD-, CAT- and GPx-like activities *via* the redox switch between $Mn^{3+}$ and $Mn^{4+}$ with a maximum reaction rate reaching 6-125 mM/min at nanomolar levels, which is unfortunately still inferior to natural enzymes.[25] Thus, the development of catalytic artificial enzymes with exceptional activity, adequate selectivity and satisfactory stability remains a major challenge for any foreseeable practical applications.

As is well known to all, most brain injuries involve enzyme-related catalytic processes and continuous neuroinflammation.[26-28] However, it is largely unclear yet which specific catalytic route(s) can be selectively targeted to inhibit neuroinflammatory responses, primarily because brain injuries simultaneously trigger various kinds of multi-enzymatic reactions between free radicals and numerous bioactive molecules.[29] Therefore, exploration of versatile artificial enzymes with different catalytic routes and desirable selectivity is beneficial to establish the relationship between oxidative stress and inflammation, and to reveal the underlying molecular pathways of catalysis.[30-32]

Atomic-level catalysts suffice a viable solution for the unmet need of improved catalytic activity and precisely modulated selectivity in a controllable manner, with lots of Fe- and Pt-based single-atom nanozymes developed.[33-40] In particular, Au contains excessive transition metal electronic states and rich electronic energy levels, which provide a solid basis for designing atomic-scale enzyme. Nevertheless, hindered by uncontrollable syntheses and complicated spatial coordination, it is difficult to reveal their electronic structures accurately, which can further influence the catalytic activity and prevent researchers from understanding exact catalytic mechanisms at atomic levels.[41] Herein, an exemplified Au-based clusterzyme was rationally designed at atomic precision with ultrahigh catalytic activity and superiority over natural antioxidants, and favorable enzymatic selectivity can be achieved *via* exquisite single-atom substitution by modulating single Cu or Cd active site, consequently serving as a promising artificial enzyme with tuned catalytic selectivity for treatment of neuroinflammation in brain.

## 2. Results

### 2.1 Structural properties of clusterzymes

Exceedingly different from most previously reported nanozymes,[1,4,5] the as-developed MPA-protected $Au_{25}$ clusterzyme is stringently defined by its unambiguous atomic configuration and geometry structure (**Figure 1a**). The hydrodynamic size of $Au_{25}$ is determined to be 2.0 nm by dynamic light scattering (DLS), and the zeta potentials of all clusterzymes are around -35 mV, suggesting the ultrasmall size and good colloid stability (**Figure S1**). The characteristic absorption at 450 nm and 670 nm of $Au_{25}$ is attributed to its unique interband transitions,[42,43] while a single-atom substitution of Cu or Cd induces a 2-3 nm minor shift, showing insignificant influence on optical properties (**Figure 1b, S2**). Electrospray ionization-mass spectra (ESI-MS) reveal a distinct *m/z* peak at ~2,271, assigned to $[Au_{25}MPA_{18}-3H]^{3-}$ (**Figure 1c**). After one atom substitution by Cu and Cd, the characteristic *m/z* peak shifts to ~2226.6 and ~2243, respectively. The inductively coupled plasma-mass spectrometry (ICP-MS) confirms that the ratios of Cd and Cu to the total metal are 5% and 4%, respectively, further validating the successful introduction of single atoms (**Figure S3**). X-ray photoelectron spectroscopy (XPS) further confirms that Au (0) is the dominant state in all clusterzymes (**Figure S4**). To identify the precise spatial atomic configuration, Extended X-ray Absorption Fine Structure (EXAFS) spectra at the Au, Cu and Cd edges were recorded (**Figure 1d**, **e** and **S5-S6**). The $L_3$ edges of Au in all clusterzymes have higher white-line intensities than the bulk standard Au foil. This is ascribed to larger surface area and alloying effects from partial oxidation with more *d*-band vacancies from nanoscale sizes and surface molecule-like interactions (Au(I)−thiolate). The characteristic absorption edges of Au clusterzymes were found at ~11920 eV, which is assigned to the 2p→5d electronic transition of Au suggesting a reduced population of unoccupied valence *d*-states. The increase of intensity in MPA-protected $Au_{24}Cu_1$ and $Au_{24}Cd_1$ indicates that the density of 5*d* electrons of Au is decreased by the one atom substitutions of Cu and Cd through the transfer of their 4*d* electrons (**Figure S5**).[44,45] The k-space oscillations of $Au_{25}$ clusters and the Au foil are shown in **Figure S5**. The k-space of the Au foil exists in typical fcc oscillation patterns which are apparently absent in all Au clusterzymes due to their small core sizes. Besides, we also investigated the XANES spectra of Cu and Cd foils as well as the corresponding atomic counterparts within clusterzymes, clearly displaying differences between single atoms and bulk metals (**Figure S6**). To further pinpoint the doping sites of Cu and Cd atoms, we performed fitting analysis on the EXAFS data of Cu and Cd. **Figure 1d** shows the R space of the EXAFS data of the Cu K edge in $Au_{24}Cu_1$. It can be seen that there is only one major peak in the range of 1.6-5.0 Å. This peak roughly corresponds to the scattering path of photoelectron waves from the X-ray absorbing Cu atom to the neighboring S atoms of different shells, and IFEFFIT program is used to fit this peak. The EXAFS parameters obtained after fitting are shown in **Table S1**. The Cu-S coordination number obtained from the fitting is 1.9 ± 0.2 Å. This value is close to 2 Å, which may indicate that the replacement of $Au_{25}$ by a Cu atom occurs at the oligomer site, consistent with previous work.[44] Similarly, the R space of EXAFS data on Cd $L_3$ edge in $Au_{24}Cd_1$ shows a peak in the range of 1.6-5.0 Å, and the fitted Cd-S coordination number is 2.3 ± 1.7 Å, which is close to the coordination number of the bond with S at the oligomer site of $Au_{25}$, indicating that Cd atom substitution may occurs at the oligomer site. (**Figure 1e**).[46]

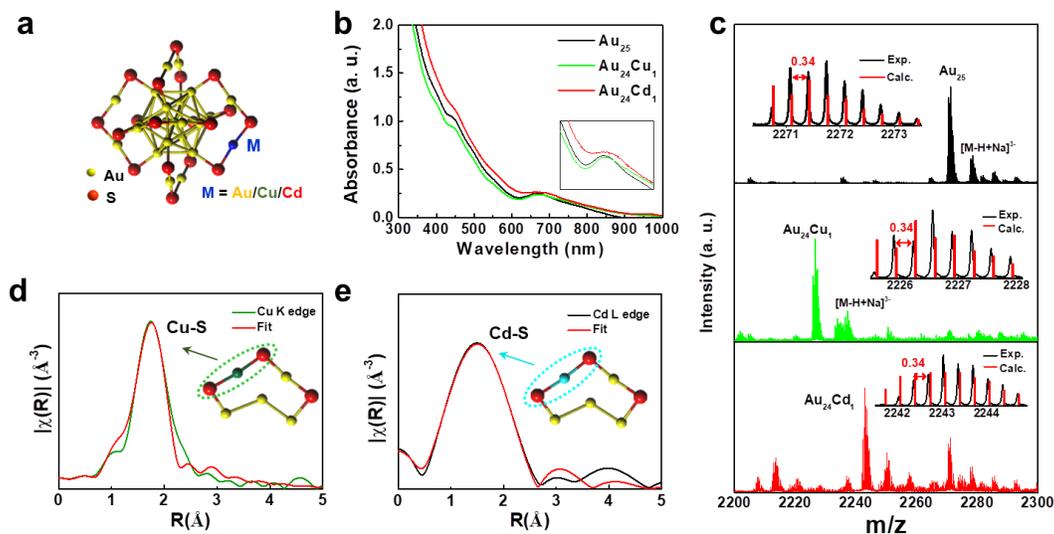

**Figure 1. Structural characterization of $Au_{25}$ clusterzymes. a** Structure illustration of $Au_{25}$, Cu and Cd substituted $Au_{24}Cu_1$ and $Au_{24}Cd_1$. **b** UV-vis absorption spectra and **c** Electrospray

ionization mass spectra (in negative ion mode) of $Au_{25}$ (black curve) before and after Cu (green curve) and Cd (red curve) substitutions. The inset in **b** is a magnification of the absorption spectrum at ~670 nm. It can be seen that the characteristic absorption is slightly redshifted after introduction of Cu and Cd, indicating decreased band gaps. The red line of the insets in **c** represents the simulated isotope distribution of $[Au_{25}MPA_{18}-3H]^{3-}$, $[Au_{24}Cu_1MPA_{18}-3H]^{3-}$, $[Au_{24}Cd_1MPA_{18}-3H]^{3-}$, respectively. **d** Cu K-edge and **e** Cd $L_3$-edge FT-EXAFS spectra and associated fitting in R space of $Au_{24}Cu_1$ and $Au_{24}Cd_1$, showing the surrounding atoms adjacent to the Cu and Cd atoms.

## 2.2 Antioxidant properties of clusterzymes

We tested the general antioxidative properties of all clusterzymes using the ABTS method (**Figure 2a**) with reference to standard antioxidants trolox and anthocyanin. Negligible antioxidant activity was observed for pure MPA-protected $Au_{25}$ clusterzyme. Single atom substitutions with Cu or Cd in the structure, however, induce a dramatic increase in antioxidant activity with increasing concentrations (**Figure 2b**). Among a variety of metals including Ag, Cu, Zn, Er, Pt, and Cd, single-atom substituted candidates $Au_{24}Cu_1$ and $Au_{24}Cd_1$ show the highest activity, representing the optimal substituting elements and ratios (**Figure S7**). Time-course kinetics of $Au_{24}Cu_1$ and $Au_{24}Cd_1$ exhibit rapid responses to the substrate in seconds with high reaction rates (**Figure 2c**). The quantitative results show $Au_{24}Cu_1$ and $Au_{24}Cd_1$ are 41 and 48 times higher in antioxidant activity than $Au_{25}$, respectively (**Figure 2c**). Compared with standard natural antioxidant controls, $Au_{24}Cu_1$ and $Au_{24}Cd_1$ are 137 and 160 times higher in activities than trolox, and 7.5 and 9 times higher than anthocyanin, respectively. The reaction rates of $Au_{24}Cu_1$ and $Au_{24}Cd_1$ at 10 and 14 μM/s are 8-11 times higher than $Au_{25}$, or 38 and 51 fold higher than trolox, respectively. In addtion, in a parellel comparison with other elements, substitution with exactly one atom of Cu or Cd present the foremost activity amidst all substituents (**Figure S7d**).

Preceding studies have evidenced that atomically precise gold clusters, such as $Au_{25}$ and $Au_{38}$, are endowed with the oxidation catalytic activities,[47-50] but their antioxidant activities are rarely reported. Herein, we discovered its ultrahigh antioxidant activity with fast kinetics *via* atom substitution.

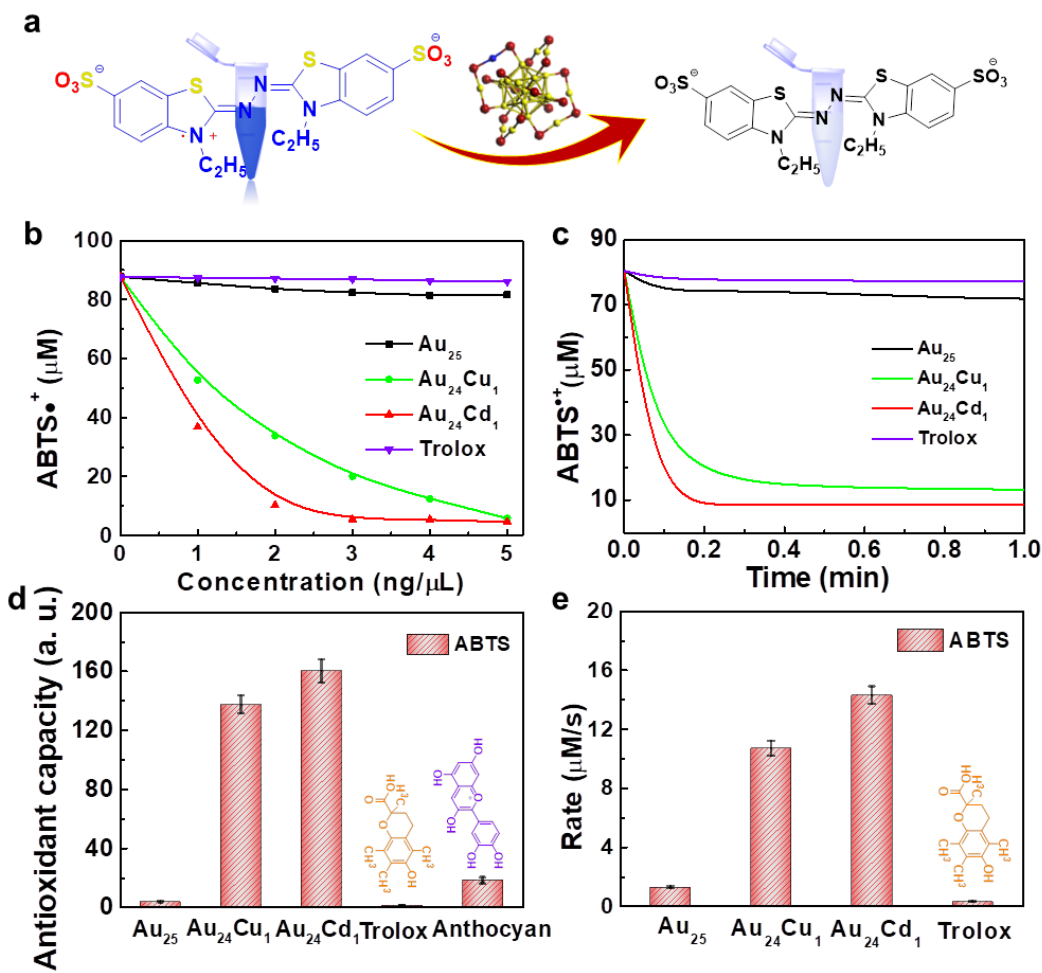

**Figure 2. The total antioxidant capacity of clusterzymes. a** Schematic illustration of the reaction mechanism of the ABTS assay. **b** Concentration- (0-5 ng/µL) and **c** time- dependent investigation of $ABTS^{•+}$ in the presence of $Au_{25}$, $Au_{24}Cu_1$ and $Au_{24}Cd_1$. **d** Comparison of the antioxidant capacities and **e** reaction rates of $Au_{25}$, $Au_{24}Cu_1$ and $Au_{24}Cd_1$ and natural antioxidants show that the antioxidant performance is greatly improved after one atom substitution with Cu or Cd. Compared with natural antioxidants, $Au_{24}Cu_1$ and $Au_{24}Cd_1$ show 137 and 160 times higher activity than trolox, and 7.5 and 9 times higher activity than anthocyanin, respectively.

## 2.3 Enzyme-like properties of clusterzymes

The general catalytic profile of clusterzymes and the schematic diagram showing catalytic processes are displayed as in **Figure 3a** and **3b**. To pinpoint catalytic selectivity of these clusterzymes, we firstly investigated the GPx-like activity of $Au_{25}$, $Au_{24}Cu_1$ and $Au_{24}Cd_1$ at the concentration of 10 ng/μL. Surprisingly, $Au_{25}$ shows the strongest tendency towards GPx-like activity with a maximum reation rate of 0.47 mM/min, higher than 0.34 mM/min for $Au_{24}Cu_1$ and 0.10 mM/min for $Au_{24}Cd_1$ (**Figure 3c**), and also significantly higher than those of previously-reported $Mn_3O_4$ nanoflowers (0.056 mM/min)[51] and Co/PMCS (0.013 mM/min).[52] The turnover frequency (TOF) value of $Au_{25}$ calculated by the Michaelis-Menten equation is 320 $min^{-1}$, 4.7 times higher than $Au_{24}Cd_1$ (**Figure S8**). This result is interesting because metals are generally considered to have low GPx-like activity, but the high GPx-like activity of $Au_{25}$ can be exploited to eliminate lipid peroxides and oxidative damages. The CAT-like activity of clusterzymes were studied at the concentration of 20 ng/μL as in **Figure 3d**. The maximum reaction rate of $Au_{25}$ is 0.074 mM/min, whereas the introduction of a Cu single atom gives rise to a 4.7-fold increase to 0.35 mM/min, suggesting its CAT-like catalytic preference. The calculated TOF value of $Au_{24}Cu_1$ for CAT-like activity is 116.7 $min^{-1}$ (**Figure S9)**, which is significantly higher than that of Pd octahedrons (1.51 $min^{-1}$).[53] The SOD-like activity of pure $Au_{25}$ can only inhibit 31 % of the substrate, while one Cd atom substitution considerably increases the inhibition rate to 89 %, empowering SOD-like selectivity (**Figure 3e**). The aforementioned results suggest enzyme-mimicking preferences of each individual clusterzyme: $Au_{25}$ as GPx, $Au_{24}Cu_1$ and $Au_{24}Cd_1$ as CAT and SOD, respectively. The structures of clustezymes before and after reaction with $H_2O_2$ suggest unchanged structures of the clusterzymes (**Figure S10-S11**).[54,55] Previous work mainly focused on the atomic substitutions of $Au_{25}$ using noble metals for catalytic reactions of hydrogen and CO/CO$_2$.[56-59] Our work herein constructively hypothesized and demonstrated that Au$_{25}$ can possess various unique enzyme-like activity modulated by single-atom substitution with non-precious metals like Cu and Cd, instead of Pt, in the geometric structure.

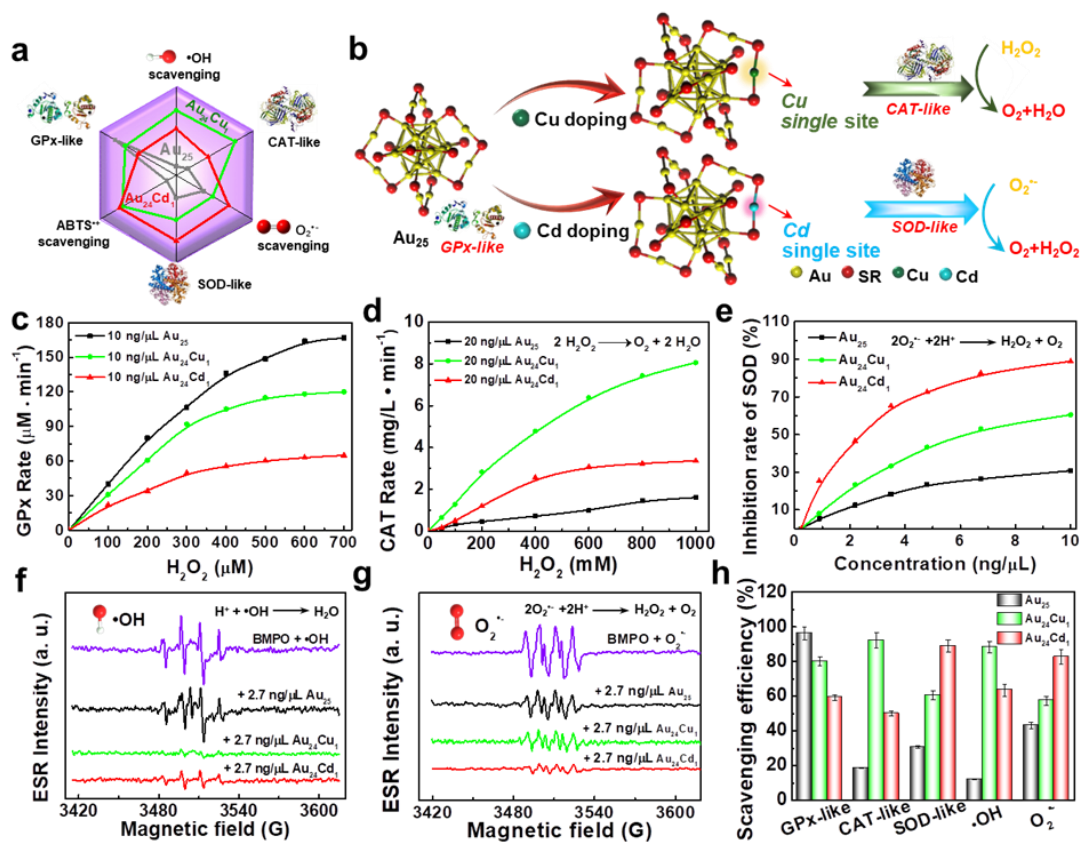

**Figure 3. Enzyme-mimetic properties and ROS scavenging activity of Au$_{25}$ clusterzymes. a** The radar map of enzymatic activities and free radical scavenging abilities of various clusterzymes. **b** Schematic illustration of catalytic selectivity of the clusterzyme system. Au$_{25}$ exhibits significant superiority in GPx-like activity; Au$_{24}$Cu$_1$ shows advantages in the CAT-like activity through the Cu single active site; Au$_{24}$Cd$_1$ preferably exhibits the SOD-like activity *via* the Cd single active site, each demonstrating a unique catalytic selectivity. **c** GPx-, **d** CAT-, **e** SOD-like activities of Au$_{25}$, Au$_{24}$Cu$_1$ and Au$_{24}$Cd$_1$. ROS scavenging activities of Au$_{25}$, Au$_{24}$Cu$_1$ and Au$_{24}$Cd$_1$ clusterzymes for **f** •OH, **g** O$_2^{•-}$ studied by the ESR spectroscopy. BMPO is used as the ROS capturing agent and the sources of •OH and $O_2^{•-}$ are $H_2O_2$ and $KO_2$, respectively. **h** Corresponding quantifications of the scavenging efficiencies.

The corresponding specific scavenging of free radicals by the clusterzymes was further investigated. The scavenging of •OH free radical was investigated using electron spin resonance (ESR) by employing BMPO as the trapping agent. The ESR signal of •OH is strong for the BMPO control, suggesting the presence of excessive •OH, while there is only a minor decrease after adding $Au_{25}$, indicative of a weak scavenging efficiency for •OH (**Figure 3f**). However, $Au_{24}Cu_1$ almost completely diminishes all ESR signals (~100 %), consistent with the observed best CAT-like activity as in **Figure 3d**. Similarly, the scavenging of $O_2^{•-}$ by clusterzymes was also investigated (**Figure 3g**). The ESR signal stays strong for the control, and slightly decreases after addition of $Au_{25}$ and $Au_{24}Cu_1$, with surplus remaining residues. In contrast, the ESR signal of $O_2^{•-}$ almost disappears in the presence of $Au_{24}Cd_1$ further validating its superior specialized SOD-like activity (**Figure 3e**). Besides, we also tested the free-radical scavenging capability of the clusterzymes towards reactive nitrogen species (RNS) including •NO, $ONOO^-$, and DPPH•. $Au_{24}Cd_1$ shows the most robust overall scavenging efficiency against DPPH• (**Figure S12**). The ESR reveals that $Au_{24}Cd_1$ has the best scavenging capability towards •NO at a low concentration of 2.7 ng/μL, whereas $Au_{25}$ presents ignorable activity (**Figure S13**). Likewise, both $Au_{24}Cd_1$ and $Au_{24}Cu_1$ also manifest significantly higher scavenging efficiency towards $ONOO^-$ than $Au_{25}$ (**Figure S14-S15**). $Au_{24}Cd_1$ is more selective against RNS than $Au_{24}Cu_1$, while $Au_{25}$ has insignificant catalytic activity. Thus, it is rational to conclude that the high selectivity for enzymes and radicals originates from the single-atom substitutions of Cu and Cd which induce redistribution of surface electrons and exert influence on electronic structures and states.

**2.4 DFT calculations and the mechanism of catalytic selectivity**

To reveal the catalytic mechanism, the Density Functional Theory (DFT) was employed to investigate the catalytic selectivity and quantum properties. By exploring possible structures in the literature, we adopt the gold core of the well-known $Au_{25}$ clusters[60] protected by ligands, which are still connected to the core *via* S atoms. To evaluate the catalytic behavior and the intermediate states during the chemical reactions, each ligand unit -$SCH_2CH_2COOH$ is simplified to -$SCH_3$. DFT optimization confirms the stability of the modeled cluster.

Due to the symmetry of the gold cluster, all possible replacements of the guest metallic atoms fall into 3 categories as follows: oligomer, the surface of core and core replacements. **Figure 4a** demonstrates the surface sites of the $Au_{13}$ core and the oligomer site replacement. The oligomer replacement is the common form which is extensively discussed in the literature, but DFT simulations indicate that the surface replacement may be another possibility. However, based on the coordination analysis in the experiment, the oligomer replacement matches the EXAFS results which yield a significantly lower coordination number (CN) than the surface replacement. With the optimized structure of the clusters, the associated CNs can be theoretically generated even within the clusters involved in the intermediate structures during the catalytic process. The averaged simulated CN values agree with the experimental values that confirm the oligomer replacement (**Table S2**). Although more theoretical investigations on the surface replacements can be found in the appendix (**Figure S16-S19**, and **Table S3**), we focus on the oligomer replacement and the associated catalytic efficiency.

Unlike the surface replacement which may cause the expansion of the core, the oligomer replacement causes the oligomer bending. It is different from the normal S-Au-S chain which aligns in a (nearly) straight line (**Figure S20**). The doped Cu shrinks the S-X-S chain while the doped Cd extends it. Compared with the typical bond length of S-Au at 2.3 Å, S-Cu and S-Cd bonds are 2.2 and 2.55 Å respectively, as shown in **Figure 4b** and **Figure S21**. With the bent chain, the distances between Cu/Cd atoms to the surface of the core are comparable, around 3.1 Å. The similarity between the Cu and Au atoms guarantees that the binding of S-Cu-S is so "firm" that the relative positions of Cu to S atoms can be hardly changed by the dynamics during the catalytic procedures which are discussed extensively below. In contrast, the relative position of the doped Cd atom may be significantly affected by the local environment such as the adhesion of small chemical units (**Figure S22-S23**).

We observed the outstanding performance of the clusterzymes in both CAT and SOD reactions with the reaction pathways summarized in **Figure 4c**. The CAT reaction usually refers to the catalytic degradation of hydrogen peroxide, and the decomposition mechanism of $H_2O_2$ may involve multiple chemical stages (**Figure 4d, 4e**). For the process of SOD, we assumed the clusters were involved in similar mechanisms to the general catalytic scheme of SOD reaction. It is worth noting that the release of oxygen completes the CAT process, while the SOD process occurs simultaneously, and the two processes are mutually permeated. The reduced cluster, Cluster(I), may also be involved in both CAT and SOD processes which depends on the concentration of different components.

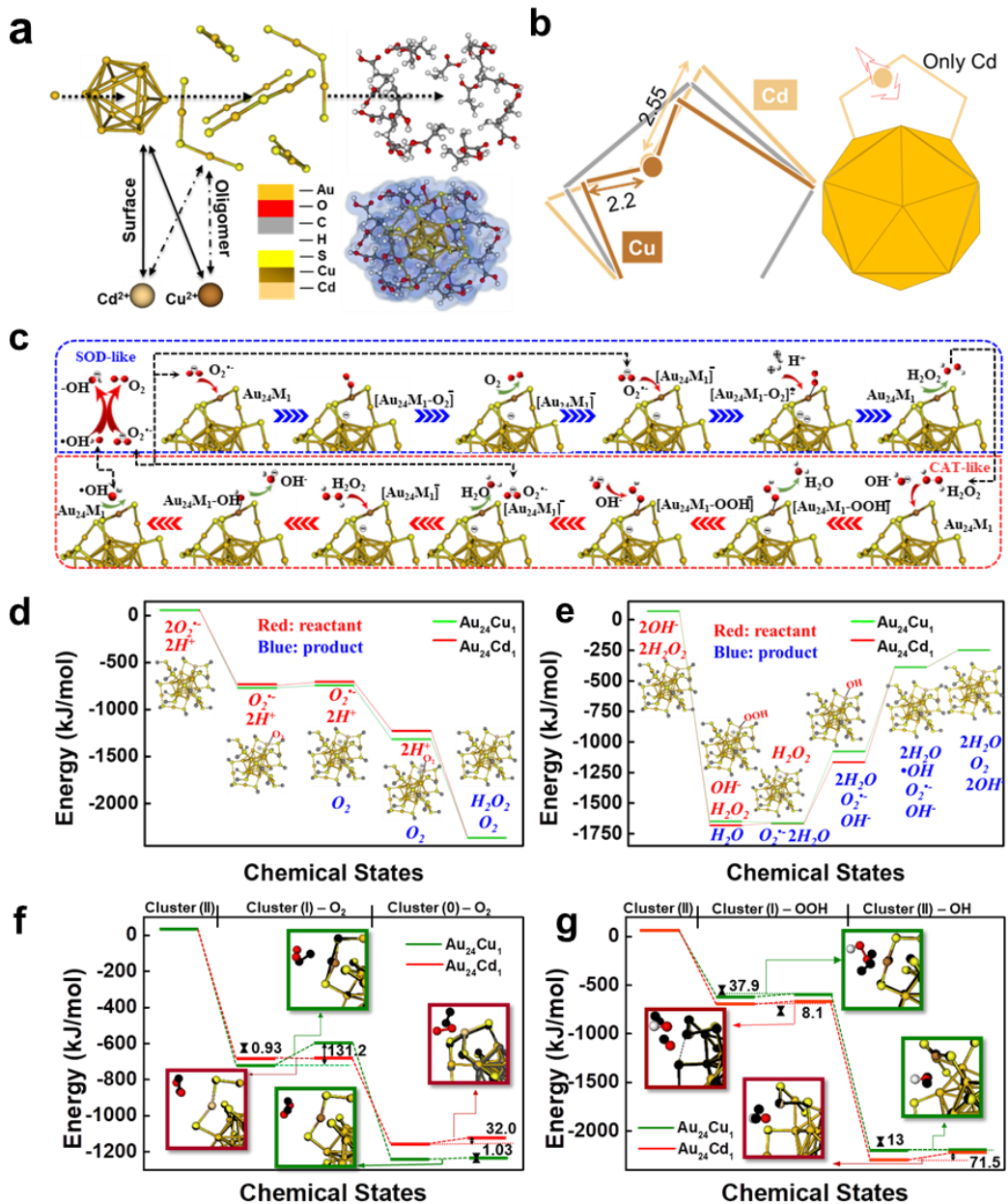

**Figure 4. DFT calculations and the mechanism of catalytic selectivity. a** Demonstration of atomic doping: surface and oligomer replacement. **b** Doped atom caused change in bonds: the more rigid Cu-S bond and the more flexible Cd-S bond. The flexibility of Cd allows the angular motion. **c** Mechanism and **d-e** energies profiles of catalytic process of the SOD and CAT processes. The black dotted line indicates that the catalytic products may be connected to other processes. Energy profiles and geometry structures of the intermediate states of **f** SOD and **g** CAT process in the lower panel. Align optimized intermediate structures (normal color) and transition structures (black).

Inspired by the Arrhenius equation, we performed the search of transition states and ground states of various types of molecules and ions to estimate the activation energies and evaluated the catalytic efficiencies. The energy profiles in **Figure 4f, 4g** agree with the behaviors of the clusterzymes in our experiments. In a series of reactions with multiple steps, the reaction rate is dominated by the slowest step, *i.e.* the transition with the largest activation energy. Such a feature can be seen in the first part of the catalytic SOD process by $Au_{24}Cu_1$. The ground state of the electrons in the corresponding intermediate structure is a triplet state suggested by DFT simulations. It indicates that the high activation energy of 131.2 kJ/mol is related to the spin matching issues which can be selective to the spin of superoxide ions. In the CAT processes, the critical step is related to the decomposition of $(cluster…OOH)^{2+}$, in which the $Au_{24}Cd_1$ exhibits higher activation energy (71.3 kJ/mol) which reduces the efficiency. The simulations clearly explain the SOD-CAT selective behaviors of the doped clusterzymes. The details of the reaction pathways are provided in the supporting information.

Our results of DFT simulations show some insights of the catalytic mechanisms. Assume in typical clusters, a substituted atom may turn into an active site itself to be involved in the catalytic process which may be accompanied by changes in the geometry. Herein, we named two mechanisms as SA (simple adhesion) and MA (bond modulated adhesion) correspondingly. The distances between the adsorbed molecule/ion and metal atoms designate the roles of the substituted atom.

The SA mechanism is mainly seen in $Au_{24}Cu_1$. Due to the firmness of the S-Cu-S oligomer, the Cu atom is relatively rigid (**Figure 4f, 4g**). The SA mechanism is also seen in the first step of SOD process catalyzed by $Au_{24}Cd_1$. The catalytic process includes the distance change and orientation change of small units. Significant changes in the distance between the active site (doped atom) and the small units are seen in most of the SOD processes. In contrast, the orientation change is the main character in most of the CAT processes.

The MA mechanism is seen in $Au_{24}Cd_1$ on the 2$^{nd}$ stage (Cluster(I)) of the SOD processes and most of the CAT processes. The bond modulation refers to the position change of the Cd atom which may deviate from the oligomer plane until a third S atom from another oligomer stops it. Thus, the S-Cd bonds are changed significantly (**Figure 4f, 4g**). The characteristics of transition and intermediate states involve the rotation of the superoxide ion (or oxygen molecule) and the position adjustment of the doped Cd atom. To be more precise, the motion of the Cd atom is along the perpendicular direction of the oligomer plane. Once a small unit joins the doped cluster to form an intermediate structure, the Cd atom sometimes leaves the oligomer plane. Therefore, the CN value of the Cd atom is larger when the Cd atom becomes the neighbor of 3 S atoms. When the Cd atom starts from its original state (S-Cd = 2.55 and 2.55 Å), passes its transition state (S-Cd = 2.63 and 2.85 Å) and arrives at the intermediate state (S-Cd = 2.57 and 2.60 Å), the angular motion is terminated (**Figure S24**). During such a process, the distance between the attached oxygen atoms is slightly expanded towards the normal distance of oxygen molecules which indicates the completion of the entire catalytic procedure (**Table S4**). A similar procedure for the $Au_{24}Cd_1$ can be observed at the adhesion of OOH- at the first stage of CAT process. Such a unique process allows the doped Cd atom to be an active site that can be self-modulated in a wide spatial range compare to the firm Cu atom. This may explain its good performance in the SOD process.

**2.5 Modulation of neuroinflammation**

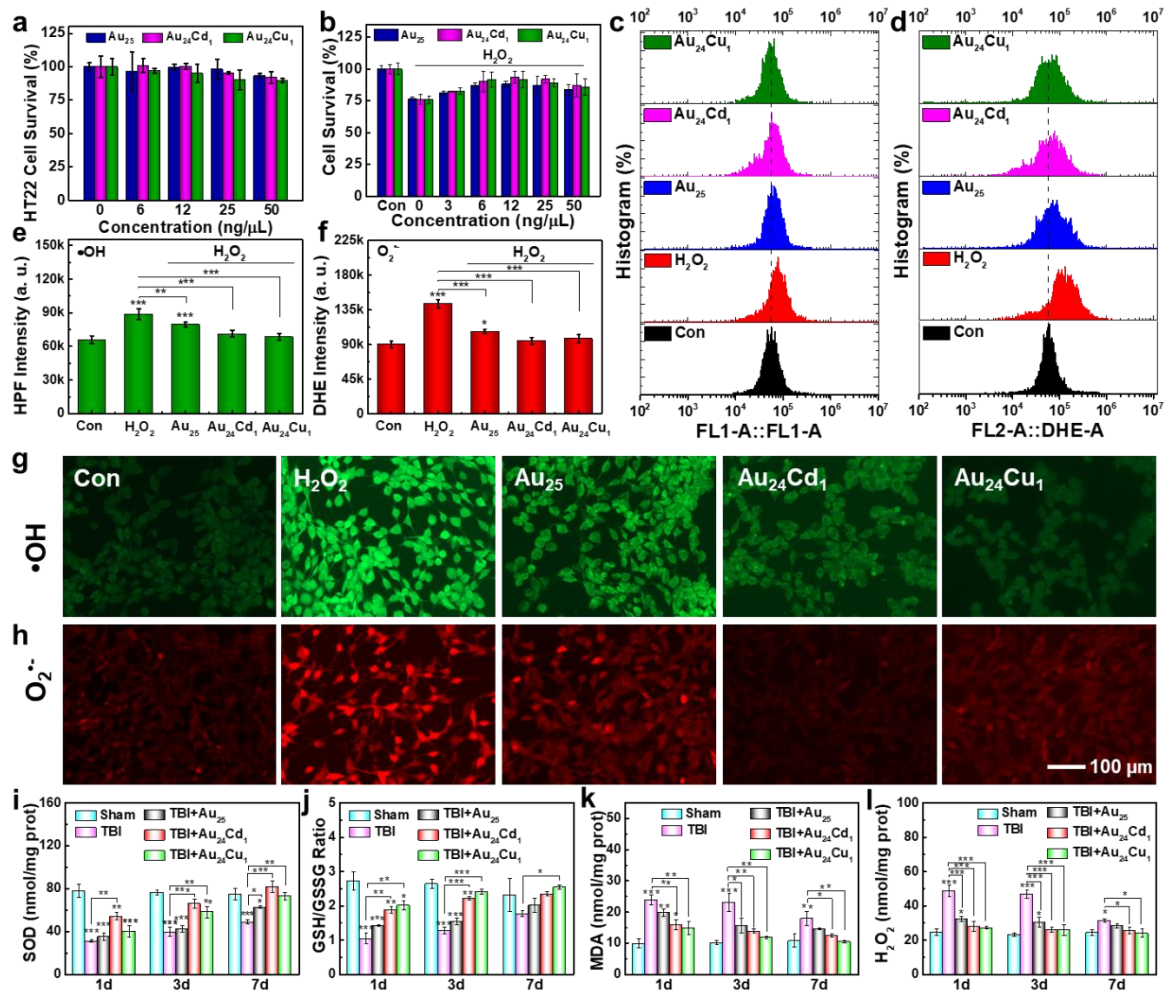

**Figure 5. Oxidative stress levels *in vitro* and *in vivo* before and after treatment of clusterzymes. a** HT22 cell viability of clusterzymes. **b** HT22 cell viability in the presence of $H_2O_2$ with or without treatment of clusterzymes as determined by MTT assays. Fluorescence quantification of cell staining for **c** and **e** •OH and **d** and **f** $O_2^{•-}$ by flow cytometry. Fluorescence microscopic images of intracellular **g** •OH (green) and **h** $O_2^{•-}$ (red) levels induced by 100 μM $H_2O_2$ with or without clusterzymes treatment, stained by HPF and DHE probes, respectively. It can be seen that $Au_{24}Cu_1$ has a better scavenging ability for •OH, which $Au_{24}Cd_1$ shows better specificity for $O_2^{•-}$, suggesting their individual selectivity for •OH and $O_2^{•-}$ respectively. **i-l** Indicators for oxidative stress including SOD, GSH/GSSG, MDA and $H_2O_2$ of TBI mice with or without treatment of clusterzymes 1, 3, and 7days post injury (n=5 per group). Data are presented as mean±SEM; *$P < 0.05$, **$P < 0.01$, and ***$P < 0.001$ compared with the Sham group, analyzed by ANOVA.

To reveal the biological activity of clusterzymes, the cell toxicity for different nerve cell lines (HT22, BV2 and MA-c) were measured by the MTT assay (**Figure 5a and S25**), showing that $Au_{25}$, $Au_{24}Cu_1$, and $Au_{24}Cd_1$ present acceptable biocompatibility. Cell survival of $H_2O_2$-stimulated neuron cells was performed with the incubation of $Au_{25}$, $Au_{24}Cd_1$ or $Au_{24}Cu_1$. As shown in **Figure 5b**, the clusterzymes treatment could improve the viability of neuron cells. To explore the correlation between the oxidative stress and the neuron viability, ROS, especially •OH and $O_2^{•-}$, were quantified and detected by a FACS flow cytometer and a fluorescence microscope using hydroxyphenyl fluorescein (HPF) and dihydroethidium (DHE) fluorescence probes, respectively. The $H_2O_2$ stimulation significantly elevates the fluorescence signal, indicating the presence of excessive amount of •OH and $O_2^{•-}$ (**Figure 5c-5h**). All clusterzymes decrease the ROS signals, with $Au_{24}Cu_1$ showing the best clearance efficiency against •OH (**Figure 5c**, **5e** and **5g**) and $Au_{24}Cd_1$ displaying the best clearance capability for $O_2^{•-}$, suggesting their individual selectivity (**Figure 5d**, **5f** and **5h**). Meanwhile, mouse models of traumatic brain injury (TBI) were used to examine the *in vivo* effects of clusterzymes. As shown in **Figure 5i-5l**, the indicators of MDA, $H_2O_2$, SOD and GSH/GSSG in the TBI group are relatively severe at day 1 post injury, but are slightly alleviated 3 days post injury, and are further improved slightly 7 days post injury. Therefore, the decrease in SOD and GSH/GSSG levels from TBI can be well rescued by clusterzymes with prominent recoveries 7 days after treatment (**Figure 5i** and **j**). Comparatively, $Au_{24}Cd_1$ induce a better recovery in SOD than $Au_{24}Cu_1$, which correlates well with their *in vitro* SOD-like activity (**Figure 3**). As the byproducts of the oxidative stress, lipid peroxides and $H_2O_2$ show higher accumulations in the brain following TBI, resulting in severe oxidative damage (**Figure 5k** and **5l**). Both $Au_{24}Cu_1$ and $Au_{24}Cd_1$ significantly inhibit the production of these harmful molecules, while $Au_{25}$ barely alters the TBI-induced increase. These results are conceivable because $O_2^{\bullet-}$ is known to be continuously produced by immediate injuries at the early stage, followed with subsequent production of lipid peroxides and $H_2O_2$. With regard to $Au_{24}Cd_1$, it can recover the diminished SOD in the first place due to its high catalytic selectivity for $O_2^{\bullet-}$, and then sustain the continuously decrease of lipid peroxides and $H_2O_2$ as the secondary catalytic options. In contrast, $Au_{24}Cu_1$ is primarily prone to increase the levels of lipid peroxides and $H_2O_2$ at the early stage due to its preference for CAT-like activity and •OH, but these molecules are intermediate products at relatively low concentrations afer TBI onset, and consequently it accounts for the increasing clearance capability in the long term.

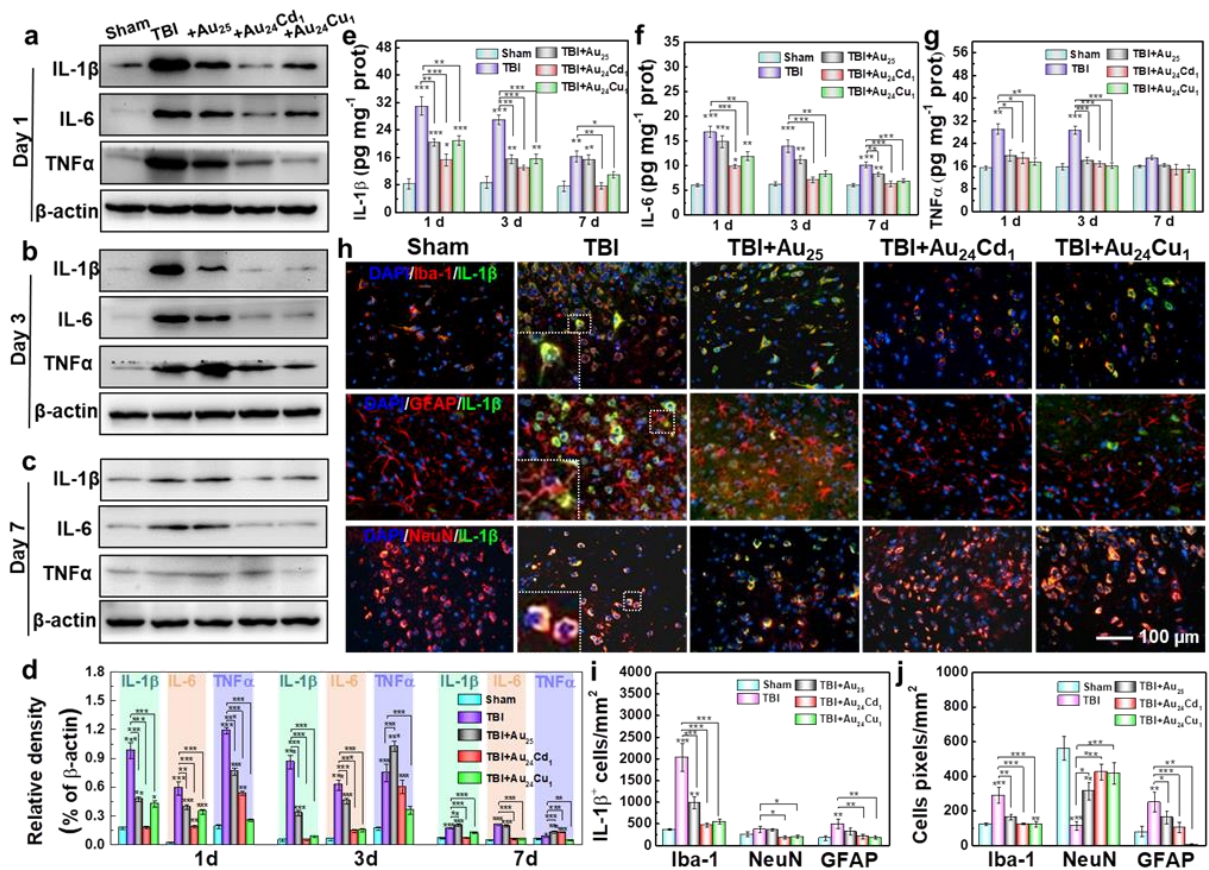

**Figure 6. Inflammation levels in brain tissues after clusterzyme treatment. a-c** Western blotting for IL-1β, IL-6, and TNFα in the brain tissues 1, 3, and 7 days post TBI after treatment (n=3 per group), respectively. **d** Western blotting quantitative analysis of inflammatory factors at different time points. It can be seen that $Au_{24}Cd_1$ can rapidly and significantly reduce the upregulated inflammatory cytokines of IL-1β and IL-6 after brain injury, while $Au_{24}Cu_1$ has a better ability to reduce the expression of TNFα. **e-g** ELISA quantitative analysis of IL-1β, IL-6, and TNFα levels in brain tissues on day 1, 3, and 7 with or without clusterzymes treatment (n=5 per group), respectively. **h** Immunofluorescence co-staining of IL-1β and microglia (Iba-1), astrocytes (GFAP) or neurons (NeuN) in injured cortex 3 days post injury with or without clusterzymes treatment. Quantitative analysis of **i** the number of IL-1β$^+$ expression in different positive cells and **j** the pixels density of Iba-1/NeuN/GFAP cells in the injured cortex with or without clusterzymes treatment (n=3 per group).Data are mean ± SEM; *$P$ < 0.05, **$P$ < 0.01, and ***$P$ < 0.001 compared with the Sham group, as analyzed by ANOVA.

Finally, the effects of clusterzymes on neuroinflammation were examined. From the Western blots and the relevant quantification analysis (**Figure 6a** and **6d**), the expression levels of IL-1β and IL-6 are significantly upregulated following TBI 1 day post injury, indicative of strong local inflammations. $Au_{24}Cd_1$ sharply downregulates IL-1β and IL-6 levels, suggesting the anti-neuroinflammation effect. In comparison, $Au_{25}$ only shows minor downregulation. Similarly, TBI leads to significant upregulation of TNFα 1 day post injury, but $Au_{24}Cu_1$ can significantly downregulate the expression of TNFα, presenting superior efficacy over $Au_{25}$ and $Au_{24}Cd_1$ (**Figure 6a**). Although the expressions of these inflammation cytokines induced by TBI are gradually suppressed by autoimmunity in the vehicle control group over time, especially 7 days post injury, there are still significant differences in IL-1β and IL-6 levels between the Sham and TBI groups (**Figure 6b-6d**). However, the clusterzymes treatment results in cytokines close to the normal level, indicating a better suppression effect on neuroinflammation. Further, the ELISA further validated the immunoblotting results that $Au_{24}Cd_1$ and $Au_{24}Cu_1$ are capable of decreasing the inflammatory cytokines in brain tissues such as IL-1β, IL-6, and TNFα, while $Au_{25}$ does not significantly alter the inflammatory cytokine patterns (**Figure 6e-6g**). $Au_{24}Cd_1$ can eliminate IL-1β and IL-6 associated inflammatory responses, while $Au_{24}Cu_1$ has a better effect on reduction of TNFα, indicating their relevant selectivity towards modulation of neuroinflammation. Finally, immunofluorescence staining of cerebral cortex harvested from mice also shows that $Au_{24}Cd_1$ and $Au_{24}Cu_1$ can remarkably decrease the TBI-elevated expressions of IL-1β, IL-6, and TNFα (**Figure 6h**, **6i and S26-S27**), therefore alleviating neuroinflammation. Colocalization studies with markers for neurons (NeuN), microglia (Iba-1) or astrocytes (GFAP) were performed in injured cortex on day 3 post injury. **Figure 6h** and **6i** reveal that IL-1β is mainly produced by microglia after TBI, similar with IL-6 and TNFα (**Figure S26-S27**). In addition, quantitative analyses of the number of positive cells show that massive microglia and astrocytes are activated and many neurons are depleted after TBI (**Figure 6j**). With the clusterzymes treatment, most of these nerve cells are rescued. Meanwhile, the morphology of TBI-activated astrocytes can be recovered to near normal levels after treatment with clusterzymes (**Figure S28**), and the neuroinflammatory responses are also prevented likewise by verified histology (**Figure S29-S32**). The clusterzymes can also restore the TBI-induced body weight loss (**Figure S33**). Moreover, behavioral tests were studied by the Morris water maze. As shown in **Figure S34a** and **S34b**, all the mice apparently learned the task during the acquisition phase of days 13-17 and 28-31, while the distance traveled and latency to hidden platform with $Au_{24}Cu_1$ and $Au_{24}Cd_1$ treatment obviously decreased. For the probe trial on day 18 and day 32 (**Figure S34c** and **S34d**), the percentage time in the missing platform quadrant and the number of platform crossingswere significantly reduced in the TBI group, but almost return to the normal level after $Au_{24}Cu_1$ and $Au_{24}Cd_1$ treatment. These results reveal trends in the improvements of learning ability and spatial memory with $Au_{24}Cu_1$ and $Au_{24}Cd_1$ treatment. In addition, we systematically studied the pharmacokinetics and toxicology of clusterzymes. It can be seen that the clusterzymes accumulated in major organs can be removed by the kidney (urine) and liver (feces). After 48 hours, ～80% of the total dose can be excreted, and most of it is excreted through the kidney (more than 70%) (**Figure S35**). No significant changes in organs or blood chemistry or hematology are found, suggesting that renal clearable clusterzymes do not cause significant biological toxicity *in vivo* (**Figure S36-S38**). Artificial enzymes have persistently been shown to exhibit multiple enzyme-like catalytic activities with a diversified class of materials.[15] Low catalyitc activity as compared to natural enzymes, however, is one of the most noticeable disadvantages due to limited electron transfers at atomic levels.[15] The rationally-designed clusterzymes with single-atom substitutions overcome such barriers with antioxidant activity 9 times higher than that of anthocyanin which is known to be one of the most reactive antioxidant molecules in nature. Besides, unlike the structurally ambiguous traditional artificial enzymes, the definitive molecular structures of clusterzymes are accurately elucidated, allowing us to distinguish the catalytically active sites and scrutinize the electronic structures and reaction energies.[62-65] As a result, the substituting single atoms can be arranged into a specific spatial location of the clusterzyme freely, thus tuning electronic structures and affecting the catalytic activity.[66-70] Meanwhile, the interactions between host atoms (i.e. Au) and the introduced substituting atoms (i.e. Cu or Cd) can induce coupled electron states and in turn influence the catatlytic selectivity.[71] In our work, the GPx-, SOD-, and CAT-like catalytic selectivity were assigned to $Au_{25}$, $Au_{24}Cd_1$, $Au_{24}Cu_1$, respectively *via* modulated bond lengths to the active center, and thus it is conceived that such a platform of clusterzymes will generate various selectivity against different molecules. By employing the three catalytically selective clusterzymes, we sucessfully established the relationship between oxidative stress and neuroinflammation, demonstrating the importance of $O_2^{\bullet-}$ and long-term benefits in TBI. Specifically, $Au_{24}Cd_1$ can significantly mitigate the neuroinflammation *via* inhibiting IL-1β and IL-6,[29,72] while $Au_{24}Cu_1$ differentially reduces neuroinflammation by inhibiting TNFα, showing selectivity against anti-neuroinflammation. Meanwhile, due to the innate ultrasmall size of clusterzymes, it can penetrate the kidney barriers and be exceted by renal, avoiding long-term hepatotoxicity and multi-organ injuries. Therefore, the clusterzymes are presumably influential as a biomedicine, especially in the field of neuroscience.

## 3. Conclusions

In summary, we report a systemic single-atom substitution approach to fabricate artifical enzymes on the basis of MPA-protected $Au_{25}$ clusters, namely clusterzymes. The clusterzymes show the ultrahigh antioxidant activity up to 137-160 times higher than the natural trolox. Moreover, the catalytic selectivity towards GPx, CAT, SOD, and nitrogen-related signaling molecules can be fine-tuned by single-atom substitutions. DFT calculations conclude that reaction pathways are modulated by the single active site of $Au_{24}Cd_1$ and $Au_{24}Cu_1$ at bond lengths. The biological results show that $Au_{24}Cd_1$ preferentially decreases IL-1β and IL-6, while $Au_{24}Cu_1$ tends to decrease TNFα, indicative of their different selectivity for modulating alleviation of neuroinflammation.

**Acknowledgements**

This work was financially supported by the National Natural Science Foundation of China (Grant No. 91859101, 81971744, U1932107, 814717866, and 11804248), the Independent Innovation Foundation of Tianjin University, the Natural Science Foundation of Tianjin (Grant No. 18JCQNJC03200) and the NSF (Grant No. IRES 1826917).

**Author contributions**

X.-D. Z. conceived and designed the experiments. H. L. contributed to materials synthesis, H. L., S. L., and X. Y. contributed to physical and chemical measurement and Y. L. and K. V. contributed to the simulation of the theoretical calculation, and S. S., Q. X., and K. C. contributed to biological experiment. X.-D. Z., J. X., S. S., H. L., X. M., H. W., W. M., and Y. L. analyzed the data, X.-D. Z., S. S., H. L., and Y. L. prepared the manuscript. All authors discussed the results and commented the manuscript.